\documentclass[twocolumn,showpacs,aps,prl,superscriptaddress]{revtex4}

\newcommand{\BABARPubYear}    {06}
\newcommand{\BABARPubNumber}  {001}

\newcommand{\SLACPubNumber} {11719}

\usepackage{graphicx}
\usepackage{dcolumn}
\usepackage{amsmath}
\usepackage{epsfig}
\usepackage{feynmp}
\input pubboard/babarsym

\newcommand{\stwobg}{\sin(2\beta\!+\!\gamma)}
\newcommand{\rhop}{\rho^{\prime+}}
\newcommand{\mpp}{m_{\pi\pi^0}}

\def\myprl  #1 #2 #3 {\jprl{#1},\ #2 (#3)}
\def\myplb  #1 #2 #3 {\plb{#1},\ #2 (#3)}
\def\myprd  #1 #2 #3 {\jprd{#1},\ #2 (#3)}
\def\mynim  #1 #2 #3 {\nim{#1},\ #2 (#3)}
\def\mypr   #1 #2 #3 {\pr{#1},\ #2 (#3)}

\begin{document}
{\pagestyle{empty}
\begin{flushleft}
\babar-PUB-\BABARPubYear/\BABARPubNumber \\
SLAC-PUB-\SLACPubNumber \\
February 2006
\end{flushleft}
\par\vskip 0.5cm
 
\title{
{\large \bf 
Measurement of time-dependent \CP asymmetries in $\Bz{\to} D^{(*)\pm}\pi^{\mp}$ and $\Bz{\to} D^{\pm}\rho^{\mp}$ decays}
}

%
\author{B.~Aubert}
\author{R.~Barate}
\author{D.~Boutigny}
\author{F.~Couderc}
\author{Y.~Karyotakis}
\author{J.~P.~Lees}
\author{V.~Poireau}
\author{V.~Tisserand}
\author{A.~Zghiche}
\affiliation{Laboratoire de Physique des Particules, F-74941 Annecy-le-Vieux, France }
\author{E.~Grauges}
\affiliation{Universitat de Barcelona, Fac. Fisica Dept. ECM, Avda Diagonal 647, 6a planta, E-08028 Barcelona, Spain }
% ...
\collaboration{The \babar\ Collaboration}
\noaffiliation

\date{\today}

\begin{abstract}
We present updated results on time-dependent \CP\ asymmetries in fully reconstructed $\Bz{\to}D^{(*)\pm}\pi^{\mp}$ and $\Bz{\to}D^{\pm}\rho^{\mp}$ decays in 
approximately $232$ million $\Y4S{\to}\BB$ events collected with the \babar\ detector at the PEP-II asymmetric-energy $B$ factory at SLAC.
From a time-dependent maximum likelihood fit we obtain for the parameters related to the \CP violation angle $2\beta\!+\!\gamma$:
\begin{eqnarray*}
\begin{array}{rclcrcl}
a^{D\pi}         \!\!&=&\!\! -0.010 \pm 0.023  \  \pm  0.007\  \!\!&,&\!
c_{\rm lep}^{D\pi}   \!\!&=&\!\! -0.033 \pm 0.042   \  \pm 0.012\   \ ,    \\ \nonumber
a^{D^*\pi}       \!\!&=&\!\! -0.040 \pm 0.023  \  \pm  0.010\   \!\!&,&\!
c_{\rm lep}^{D^*\pi} \!\!&=&\!\! \phantom{-} 0.049 \pm 0.042 \  \pm  0.015\ \ ,    \\ \nonumber
a^{D\rho}        \!\!&=&\!\! -0.024 \pm 0.031  \  \pm  0.009\   \!\!&,&\!
c_{\rm lep}^{D\rho}  \!\!&=&\!\! -0.098\pm 0.055   \  \pm  0.018\  \ ,     \\
\nonumber
\end{array}
\end{eqnarray*}
where the first error is statistical and the second is systematic.
Using other measurements and theoretical assumptions, we interpret the results in terms of the angles of the
Cabibbo-Kobayashi-Maskawa unitarity triangle and find 
$|\stwobg|\!>\!0.64\ (0.40)$ at $68\%\ (90\%)$ confidence level.
\end{abstract}


\maketitle

In the Standard Model, \CP\ violation in the weak interactions between quarks
manifests itself as a non-zero area of the Cabibbo-Kobayashi-Maskawa (CKM) unitarity triangle~\cite{CKM}.
While the measurement of $\sin 2\beta$ is now quite precise~\cite{babar_sin2b,belle_sin2b},
the constraints on the other two angles of the unitarity triangle, $\alpha$ and $\gamma$, are
still limited by statistical and theoretical uncertainties.

This paper presents updates for the measurements of \CP\ asymmetries in $\Bz{\to} D^{(*)\pm}\pi^{\mp}$ decays~\cite{chconj},
as reported in Ref.~\cite{olds2bg}, with a larger data sample ($\times 2.6$), and in addition includes the measurement of the
\CP\ asymmetry in the decay mode $\Bz{\to} D^\pm \rho^{\mp}$.
We denote these decays as $\Bz{\to} D^{(*)\pm} h^{\mp}$, where $h^{\mp}$ is a charged pion or $\rho$ meson.

The time evolution of $\Bz{\to} D^{(*)\pm} h^{\mp}$ decays 
is sensitive to $\gamma$ because the CKM-favored decay amplitude $\Bzb{\to} D^{(*)+} h^-$, which is proportional to the CKM matrix
elements $V^{}_{cb}V^*_{ud}$, and the doubly-CKM-suppressed decay amplitude $\Bz{\to} D^{(*)+}h^-$, 
which is proportional to $V_{cd}V^*_{ub}$, interfere due  to  $\Bz\!-\!\Bzb$  mixing. 
The relative weak phase between these two amplitudes is $\gamma$. 
With  $\Bz\!-\!\Bzb$  mixing, the total weak phase difference between the interfering amplitudes is $2\beta\!+\!\gamma$.

Neglecting the very small decay width difference between the two \Bz mass eigenstates~\cite{PDG}, the proper-time distribution of the $\Bz{\to} D^{(*)\pm}h^{\mp}$
decays is given by
\begin{eqnarray}
f^{\pm}(\eta,\deltat) &=& \frac{e^{-\left|\deltat\right|/\tau}}{4\tau} \times [1 \mp S_\zeta \sin(\deltamd\deltat) \\ \nonumber 
&\mp&\eta\, C \cos(\deltamd\deltat)]\,,
\label{eq:fplus}
\end{eqnarray}
where $\tau$ is the \Bz lifetime, $\deltamd$ is the $\Bz\!-\!\Bzb$ mixing frequency,
and $\deltat = t_{\rm rec} - \t_{\rm tag}$ is the time difference between the
$\Bz{\to} D^{(*)\pm} h^{\mp}$ decay ($B_{\rm rec}$) and the decay of the other $B$ ($B_{\rm tag}$) from the \upsbzbz decay.
In this equation the upper (lower) sign refers to the flavor of $B_{\rm tag}$ as \Bz(\Bzb),
while $\eta=+1$ ($-1$) and $\zeta=+$ ($-$) for the final state $D^{(*)-}h^{+}$ ($D^{(*)+}h^{-}$).
The sine term is due to interference between direct decay and decay after $\Bz\!-\!\Bzb$ mixing. 
The cosine term arises from interference between decay amplitudes with different weak and strong phases (direct \CP violation) or from \CP violation in mixing. 
The $S$ and $C$ asymmetry parameters can be expressed as 
\begin{eqnarray}
S_\pm = -\frac{2\textrm{Im}(\lambda_\pm)}{1+|\lambda_\pm|^2}\,, \hspace{0.4cm} {\rm and}
\hspace{0.4cm}
C=\frac{1-r^2}{1+r^2}\,,
\label{eq:cands}
\end{eqnarray}
where  $r\equiv |\lambda_+| = 1/|\lambda_-|$ and
\begin{eqnarray}
\label{eq:lambda}
 \lambda_\pm =\frac{q}{p}
\frac{A(\Bzb{\to} D^{(*)\mp}h^\pm)}{A(\Bz{\to}D^{(*)\mp}h^\pm)}
=r^{\pm 1}e^{-i(2\beta+\gamma\mp\delta)}.
\end{eqnarray}
Here $\frac{q}{p}$ is a function of the elements of the mixing Hamiltonian~\cite{PDG}, and $\delta$ is the relative strong phase between the
two contributing amplitudes. 
In the Standard Model, \CP\ violation in mixing is negligible and thus $|\frac{q}{p}|=1$.
In these equations, the parameters $r$ and $\delta$ depend on the choice of the final state. 
They will be 
indicated as  $r^{D\pi}$, $\delta^{D\pi}$ for the  $\Bz{\to}D^{\pm}\pi^{\mp}$ mode, $r^{D\rho}$, $\delta^{D\rho}$ for $\Bz{\to}D^{\pm}\rho^{\mp}$,
and $r^{D^*\pi}$, $\delta^{D^*\pi}$ for $\Bz{\to}D^{*\pm}\pi^{\mp}$~\cite{fleischer,note}.

Interpreting the $S$ parameters in terms of the angles of the unitarity triangle requires knowledge of the corresponding $r$ parameters. 
The values of $r$ are expected to be small ($\sim\!\!0.02$) and therefore cannot be extracted from the measurement
of $C$. 
They can be estimated, assuming $SU(3)$ symmetry and neglecting contributions from $W$-exchange diagrams,
from the ratios of branching fractions $\BR(\Bz{\to} D_s^{(*)+}\pi^-)/{\BR(\Bz{\to}D^{(*)-}\pi^+)}$ and $\BR(\Bz{\to} D_s^{+}\rho^-)/{\BR(\Bz{\to}D^{-}\rho^+)}$~\cite{sin2bg,olds2bg,partial,dsrho}.

This measurement is based on $232$ million \upsbb decays, collected 
with the \babar\ detector~\cite{detector} at the PEP-II asymmetric-energy $B$ factory at SLAC.
We use a Monte Carlo simulation of the \babar\ detector based on
GEANT4~\cite{geant} to validate the analysis procedure and to estimate some of the backgrounds.

The event selection criteria are unchanged from our previous publication~\cite{olds2bg}, 
except for the application of a kaon veto on the pion candidate in the decay modes $D^{(*)-}\pi^+$ to suppress $\Bz{\to} D^{(*)-}K^+$ background events, 
and for the addition of the decay mode $\Bz{\to}D^{-}\rho^{+}$. 
The $D^{*-}$ is reconstructed through its decay to $\bar{D}^0\pi^{-}$, where the $\bar{D}^0$ decays into $K^{+}\pi^{-}$, $K^{+}\pi^{-}\piz$, $K^{+}\pi^{-}\pi^{+}\pi^{-}$, or $\KS\pi^{+}\pi^{-}$. 
The $D^{-}$ is reconstructed through its decay into $K^{+}\pi^{-}\pi^{-}$ or $\KS\pi^{-}$.
The $\rho^+$ decay is reconstructed in the final state $\pi^+\piz$. 
For the \CP\ analysis we require the $\pi^+\piz$ invariant mass ($\mpp$) to be in the window $620< \mpp <920$ $\rm{MeV}/c^2$. 
Exploiting the polarization of the $\rho$ meson from the decay $\Bz {\to} D^{-}\rho^{+}$ we require  the cosine of the $\rho^+$ helicity angle $\theta_{\rm hel}$, 
defined as the angle between the charged pion and the $D^-$ momentum in the $\rho^+$ rest frame, to satisfy $|\cos\theta_{\rm hel}|\!>\!0.4$.

The beam-energy substituted mass, $\mes \equiv \sqrt{s/4 - {p_B^*}^2}$, and the difference between the $B$ candidate's measured energy and the beam energy, 
$\DeltaE \equiv E_{B}^* - (\sqrt{s}/2)$, are used to 
identify the final sample, where $E_{B}^*$ ($p_B^*$) is the energy (momentum) of the \B\ candidate in the nominal $e^{+}e^{-}$ center-of-mass frame, 
and $\sqrt{s}$ is the total center-of-mass energy.
The $\DeltaE$ signal region is defined as $|\DeltaE|<3\sigma$, where the resolution $\sigma$ is mode-dependent and approximately $20~\mev$, as determined from data. 
Figure~\ref{fig:mes} shows the $\mes$ distribution for candidates with $\mes\!>\!5.2~\gevcc$ in the \DeltaE\ signal region.
These candidates satisfy the tagging and vertexing requirements, which are described later.
Each distribution is fit to the sum of a threshold function~\cite{Argus}, which accounts for the background from random combinations of tracks (combinatorial background), 
and a Gaussian distribution with a fitted width of about $2.5~\mevcc$, which describes the signal and the backgrounds that peak in the $\mes$ signal region (peaking background). 
\begin{figure}[!htb]
\begin{center}
\includegraphics[width=8.5cm]{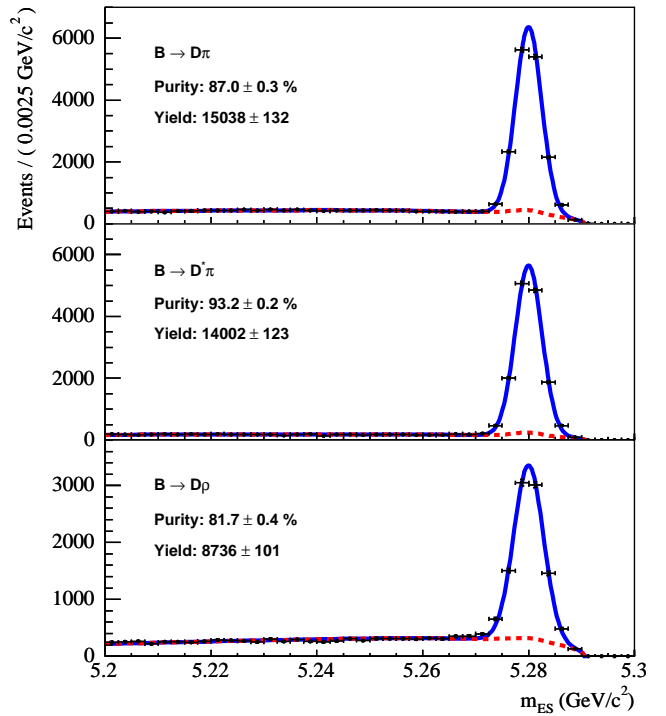}
\caption{
\mes\ distributions in the signal region for, from top to bottom, the $B{\to} D^{\pm}\pi^{\mp}$, $B {\to} D^{*\pm}\pi^{\mp}$, 
and $B {\to} D^{\pm}\rho^{\mp}$ sample for the events that satisfy the tagging and vertexing requirements described in the text, fit with the function described in the text.
The dashed lines indicate the sum of the combinatorial and peaking background contributions.}
\label{fig:mes}
\end{center}
\end{figure}
Signal yields and sample purities are determined in the $\mes$ signal region, with $\mes \!>\! 5.27\ \gevcc$, and are summarized in Table~\ref{tab:yields}.
Backgrounds from $\Bz$ and $\B^+$ decays that peak in the $\mes$ signal region are estimated using Monte Carlo events and are mostly due to charmed final states.
They are also reported in Table~\ref{tab:yields}.
\begin{table}[ht]
\begin{center}
\caption{
Signal yields, sample purities $P$, and fractions of peaking backgrounds, $f_{peak}$, for the selected samples for events that satisfy the tagging and vertexing requirements described in the text. 
} \label{tab:yields}
\begin{tabular}{|l|c|c|c|c|}\hline
Decay  & Yield & $P (\%)$ & \multicolumn{2}{c|}{$f_{peak}$(\%)} \\ 
mode &  & &  \Bz & \Bpm\\  
\hline
$B {\to}D^{\pm}\pi^{\mp}$    & $15038 \pm 132$    & $87.0 \pm 0.3$    & $1.6 \pm 0.1$    &  $1.2 \pm 0.1$  \\
$B {\to}D^{*\pm}\pi^{\mp}$   & $14002 \pm 123$    & $93.2 \pm 0.2$    & $1.0 \pm 0.1$    &  $1.1 \pm 0.1$  \\
$B {\to}D^{\pm}\rho^{\mp}$   & $8736  \pm 101$    & $81.7 \pm 0.4$    & $1.3 \pm 0.2$    &  $1.5 \pm 0.2$  \\
\hline
\end{tabular} 
\end{center}
\end{table}

For the $\Bz {\to} D^{\pm}\rho^{\mp}$ mode we consider additional sources of background with the same final state
$D^{\pm}\pi^{\mp}\piz$, where the $\pi^{\mp}\piz$ system is not produced through the $\rho^{\mp}$ resonance.
Interfering sources of background can introduce a dependence of the $\lambda^{D\rho}_{\pm}$ parameters of Eq.~\ref{eq:lambda} on $\mpp$.
The dependency has been studied using the distribution of $\mpp$.

The possible background contributions have been evaluated with a sample of $130273$ $\Bz{\to} D^-\pi^+\pi^0$ candidates, on which the requirements on the $\rho$ helicity and on $\mpp$ have been removed.
Three interfering components are considered: $\Bz {\to} D^-\rho^+$ (the signal), $\Bz {\to} D^-\rhop(1450)$  with a pole mass of $(1465\pm25)\mevcc$ and a width of $(400 \pm 60)\mevcc$~\cite{PDG} for the $\rhop$, 
both described with P-wave relativistic Breit-Wigner functions~\cite{breitwigner,bwfactor}, and a
non-resonant component, $\Bz {\to} D^-(\pi^+\piz)_{nr}$.
Contributions from the decay modes $\Bz {\to} D^{*-}\pi^+$ $(D^{*-}{\to} D^{-} \pi^0)$ and $\Bz {\to} \bar{D}^{**0}\pi^0$ $(\bar{D}^{**0}{\to} D^{-}\pi^+)$ 
are negligible due to the kinematic constraints imposed on the $\rho$ daughter particles.
We perform a fit to the binned $\mpp$ distribution to extract the amplitudes of the three components, 
where for each bin the combinatorial background has been subtracted, as estimated from the corresponding \mes\ distribution, and the number of peaking background events has been 
estimated using fully simulated Monte Carlo events.
The result of the fit is shown in Fig.~\ref{fig:mpipi0}.
The fraction of $\Bz {\to} D^-\rhop(1450)$ and $\Bz {\to} D^-(\pi^+\piz)_{nr}$ events in 
the mass window $620<\mpp<920~\rm{MeV}/c^2$ is found to be smaller than $0.02$ at $90\%$ confidence level (C.L.)
\begin{figure}[!htb]
\begin{center}
\epsfig{figure=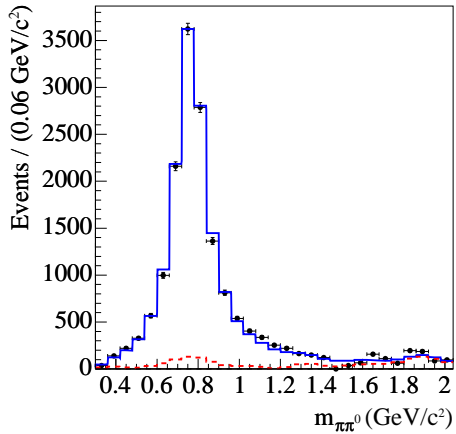,width=6.0cm,clip=}
\caption{
$\mpp$ distribution for the combinatorial-background subtracted $B {\to} D^{\pm}\pi^{\mp}\pi^0$ sample,
containing 16214 events.
The solid line is the fit projection, consisting of the three interfering components described in the text 
and an \mes\ peaking background contribution, indicated with the dashed line.
}
\label{fig:mpipi0}
\end{center}
\end{figure}

The proper time interval $\deltat$ between the two $B$ decays is calculated from the measured separation $\deltaz$, between the $B_{\rm rec}$ and $B_{\rm tag}$ decay points along the
beam direction. We determine the  $B_{\rm rec}$ decay point from its charged tracks. 
The  $B_{\rm tag}$ decay point is obtained by fitting tracks that do not belong to $B_{\rm rec}$, 
imposing constraints from the $B_{\rm rec}$ momentum and the beam-spot location. We accept events with calculated $\Delta t$ uncertainty of less than $2.5~$ps and $|\Delta t|<20~$ps.
The average $\deltat$ resolution is approximately $1.1~$ps.
We use multivariate algorithms that identify signatures in the $B_{\rm tag}$ decay products to determine (``tag'') the flavor to be either a \Bz or a \Bzb~\cite{babar_sin2b}. 
Primary leptons from semileptonic $B$ decays are selected from identified electrons and muons and from isolated energetic tracks. 
The charges of identified kaons and soft pions from $D^{*+}$ decays are also used to extract flavor information. 
Each event with an estimated mistag probability less than $45\%$ is assigned to one of six hierarchical, mutually exclusive tagging categories. 
The lepton tagging category contains events with an identified lepton, while other events are divided into categories based on their estimated mistag probability. 
The effective efficiency of the tagging algorithm, defined as $Q = \Sigma_i\, \epsilon_i(1-2w_i)^2$, where $\epsilon_i$ and $w_i$ are the efficiency 
and the mistag probability, respectively, for category $i$, is $30.1 \pm 0.5\%$.

Since the expected \CP\ asymmetry in the selected $B$ decays is small,
this measurement is sensitive to the interference between the $b{\to}u$ and  $b{\to}c$ amplitudes in the decay of $B_{\rm tag}$. 
To account for this ``tagside interference'', we use a parametrization different from Eq.~\ref{eq:fplus}, which is described in Ref.~\cite{DCSD} and summarized here. 
For each tagging category $i$, independent of the decay mode $\mu \in \{D\pi, D^{*}\pi, D\rho\}$, the tagside interference is parametrized in terms of the 
effective parameters $r^\prime_i$ and $\delta^\prime_i$.
Neglecting terms of order $(r^{\mu})^2$ and $(r^{\prime}_i)^2$, the \deltat distributions are written as
\begin{eqnarray}
f^{\pm,{\mu}}_i(\eta,\deltat) &=&
  \frac{e^{-\left|\deltat\right|/\tau}}{4\tau} \times [ 1 \mp (a^{\mu}
  \mp \eta b_i - \eta c_i^{\mu})\nonumber \\
  && \sin(\deltamd\deltat)\mp\eta\cos(\deltamd\deltat)]\,,
\label{completepdf}
\end{eqnarray}
where, in the Standard Model,
\begin{eqnarray}\nonumber
&a^{\mu}&=\ 2r^{\mu}\sin(2 \beta\!+\!\gamma)\cos\delta^{\mu}\,, \\ \nonumber
&b_i&=\ 2r^\prime_i\sin(2 \beta\!+\!\gamma)\cos\delta^\prime_i\,, \\
&c_i^{\mu}&=\ 2\cos(2 \beta\!+\!\gamma) (r^{\mu}\sin\delta^{\mu}-r^\prime_i\sin\delta^\prime_i)\,.
\label{acdep}
\end{eqnarray}
Semileptonic $B$ decays do not have a doubly-CKM-suppressed amplitude contribution, and hence $r^{\prime}_{\rm lep}=0$. 
In the following, we quote results for the six $a^\mu$ and $c^\mu_{\rm{lep}}$ parameters, which are independent of the unknown parameters $r^\prime_i$ and $\delta^\prime_i$.
The other $b_i$ and $c_i^\mu$ parameters depend on $r^\prime_i$ and $\delta^\prime_i$, 
and do not contribute to the interpretation of the result in terms of $\stwobg$. Note that all tagging categories contribute to the measurement of the $a^{\mu}$ parameters.

An unbinned maximum-likelihood fit is applied to the  \deltat\ distribution of the selected $B$ candidates in the $\Delta E$ signal region. The whole $\mes$ range is used to determine the signal probability of each event on the basis of the Argus plus Gaussian fit described previously.
The effect of finite \deltat\ resolution is described by convoluting Eq.~\ref{completepdf} with a resolution function composed of three Gaussian distributions. 
Incorrect tagging dilutes the parameters $a^{\mu}$, $c_i^{\mu}$, and the coefficient of $\cos(\deltamd\deltat)$ by a factor $D_i=1-2w_i$~\cite{babar_sin2b}.
The parameters of the resolution function and those associated with flavor tagging are determined from the fit to the data and are consistent with previous \babar\ analyses~\cite{babar_sin2b}.
The $\Delta t$ distribution of the combinatorial background is parametrized using two empirical components: a prompt component with zero lifetime and a component with an effective lifetime.  
The components are convoluted with the sum of two Gaussians, and the
resolution parameters of the two Gaussians, including the effective dilution parameters, 
the effective lifetime, and the relative fraction of the two components, are determined from the fit to the data. 
The peaking background coming from $B^{\pm}$ mesons is modeled by an exponential with the $B^{\pm}$ lifetime.
Its relative fraction is fixed to the value estimated from simulations. 
The resolution function is the same as the signal resolution, while the dilution parameters are fixed to the values obtained from a $B^+$ control sample.
The peaking backgrounds from \Bz mesons, whose amounts are also fixed to the value estimated using simulation, 
are modeled with a likelihood similar to the signal likelihood, but without \CP violation (all the $a$, $b$, $c$ parameters set to zero). 
Possible \CP violation in this background is taken into account in the evaluation of the systematic uncertainties. 
The resolution and the dilution parameters are the same as for the signal.

From the unbinned maximum likelihood fit we obtain:
\begin{eqnarray} \label{acmeas} 
a^{D\pi}              &=& -0.010\pm 0.023  \,\pm 0.007 \,\,,  \\ \nonumber
c_{\rm lep}^{D\pi}    &=& -0.033\pm 0.042  \,\pm 0.012 \,\,, \\ \nonumber
a^{D^*\pi}            &=& -0.040\pm 0.023  \,\pm 0.010 \,\,, \\ \nonumber
c_{\rm lep}^{D^*\pi}  &=& \phantom{-}0.049\pm0.042  \,\pm 0.015 \,\,,\\ \nonumber
a^{D\rho}             &=& -0.024\pm 0.031  \,\pm 0.009 \,\,, \\ \nonumber
c_{\rm lep}^{D\rho}  &=& -0.098\pm 0.055  \,\pm 0.018 \,\, ,
\end{eqnarray} 
where the first quoted error is statistical and the second is systematic.
The largest correlation with any linear combination of other fit parameters is about $20\%$ and $30\%$ for the $a^\mu$ and the $c_{\rm {lep}}^\mu$ parameters, respectively.
Figure~\ref{fig:dt1} shows the fitted $\deltat$ distributions for events from the lepton
tagging category, which has the lowest level of backgound and mistag probability.
\begin{figure}[!htb]
\begin{center}
\epsfig{figure=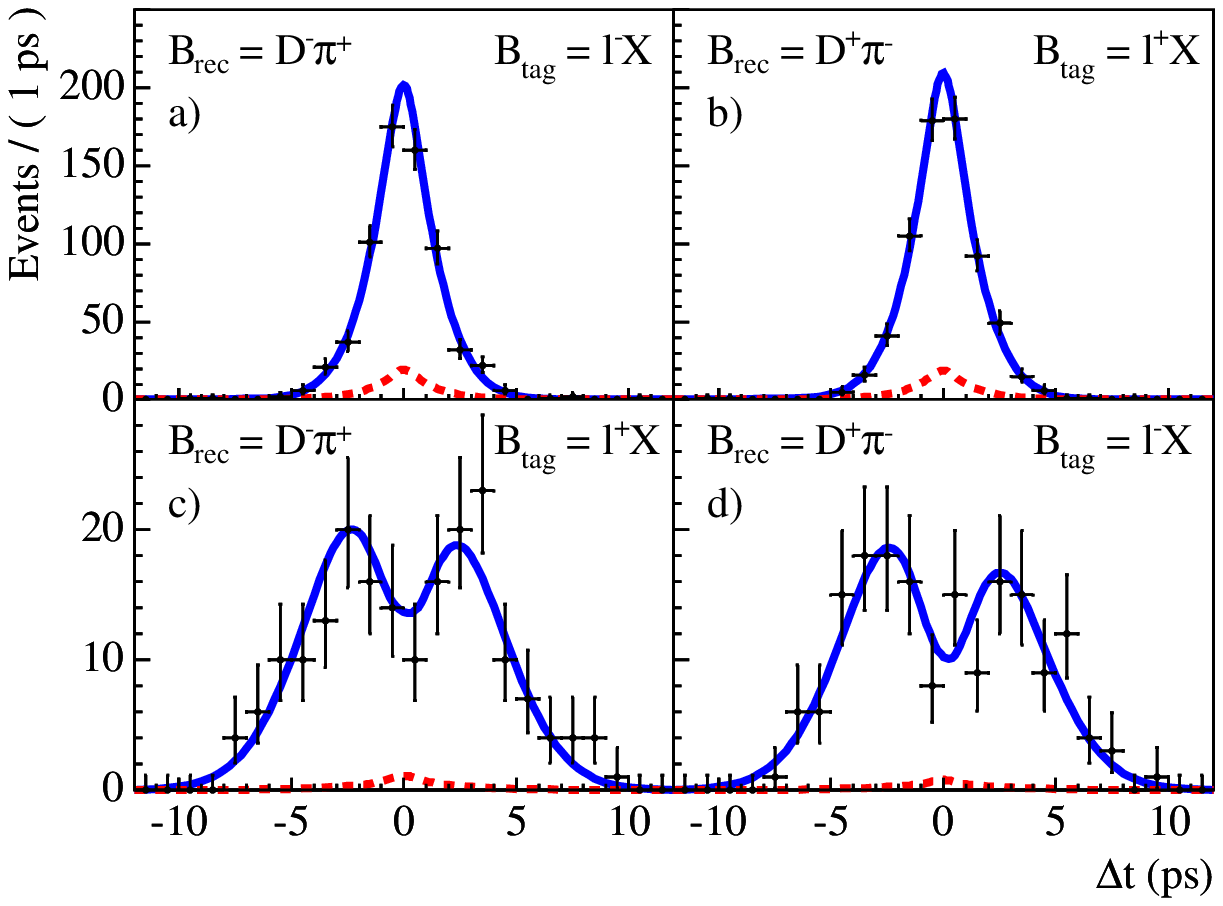,width=8.0cm,clip=}
\epsfig{figure=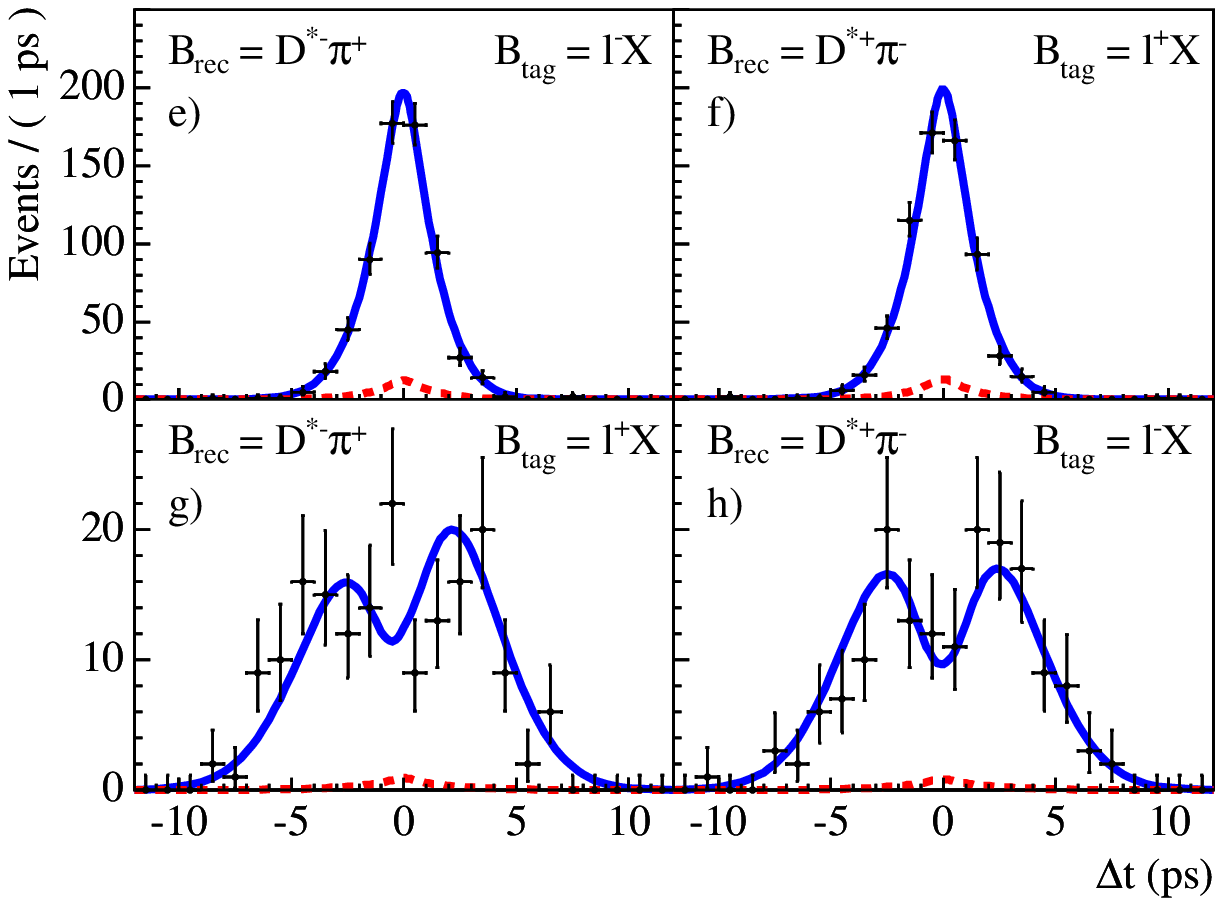,width=8.0cm,clip=}
\epsfig{figure=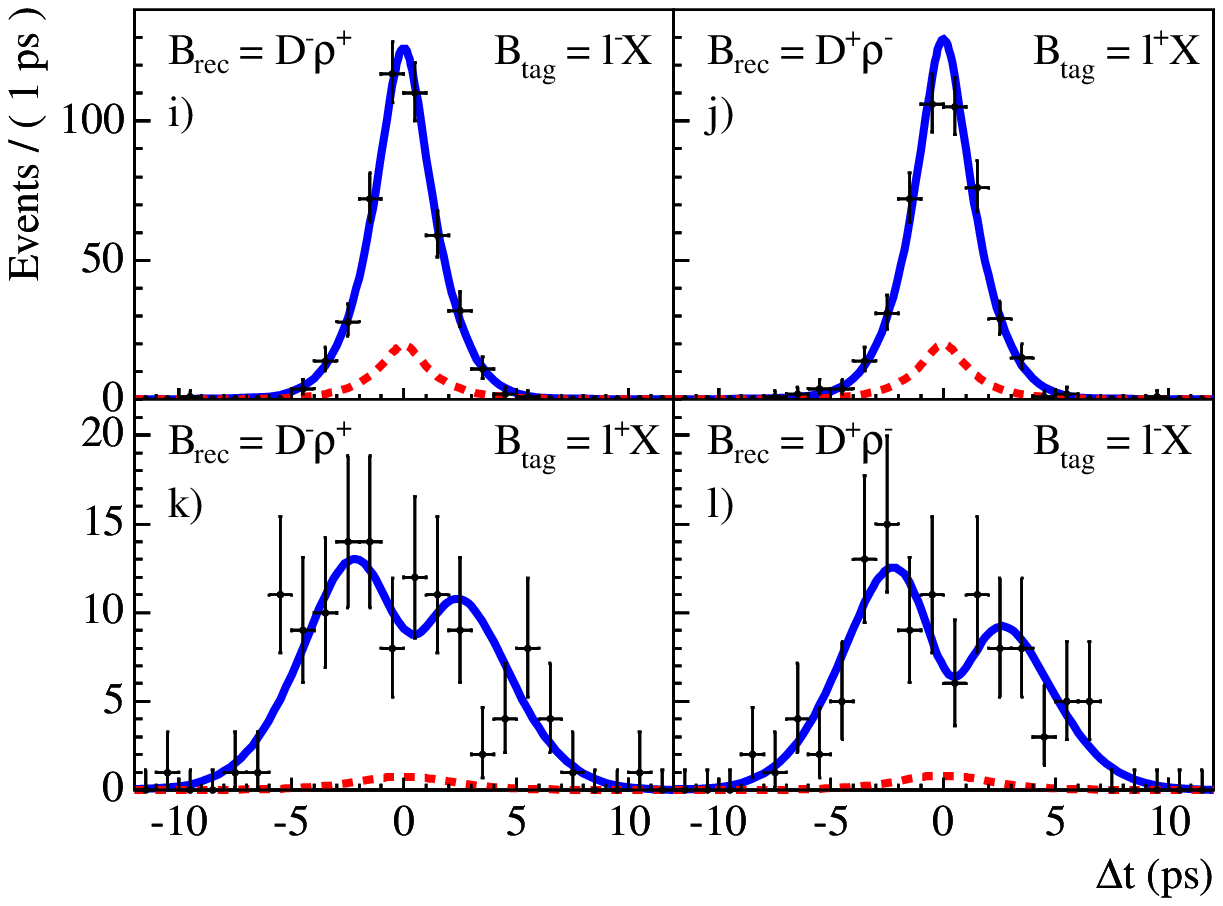,width=8.0cm,clip=}
\caption{
Distributions of \deltat\ for the  $\Bz{\to} D^{\pm}\pi^{\mp}$ (a-d), $\Bz{\to} D^{*\pm}\pi^{\mp}$ (e-h), and $\Bz{\to} D^{\pm}\rho^{\mp}$ (i-l) candidates tagged with leptons,
split by  $B$ tagging flavor and  reconstructed final state. The solid lines are fit projections. 
The  background contributions are represented by the dashed curves.
}
\label{fig:dt1}
\end{center}
\end{figure}
The various contributions to the systematic uncertainties of the $a$ and $c_{\rm{lep}}$ parameters are shown in Table~\ref{tab:sys}.
\begin{table}[ht]
\begin{center}
\caption{Systematic uncertainties on the $a$ and $c_{\rm lep}$ parameters (in units of $10^{-2})$.} \label{tab:sys}
\begin{tabular}{|l|l|l|l|l|l|l|}\hline
\Bz\ mode  & \multicolumn{2}{c}{ $D^{\pm}\pi^{\mp}$}& \multicolumn{2}{|c|}{$D^{*\pm}\pi^{\mp}$} & \multicolumn{2}{c|}{$D^{\pm}\rho^{\mp}$}\\ \hline
Source & $\sigma_{a}$ & $\sigma_{c}$ & $\sigma_{a}$ & $\sigma_{c}$ & $\sigma_{a}$ & $\sigma_{c}$  \\
\hline
Vertexing ($\sigma_{\Delta t}$)   & 0.37 & 0.64 & 0.80 & 1.14 & 0.47 & 1.15 \\
Fit ($\sigma_{\rm fit}$)          & 0.51 & 0.95 & 0.52 & 0.99 & 0.75 & 1.34 \\
Model ($\sigma_{\rm mod}$)        & 0.12 & 0.13 & 0.12 & 0.13 & 0.01 & 0.18 \\
Tagging ($\sigma_{\rm tag}$)      & 0.07 & 0.16 & 0.11 & 0.14 & 0.06 & 0.12 \\
Background ($\sigma_{\rm bkg}$)   & 0.13 & 0.10 & 0.10 & 0.09 & 0.28 & 0.29 \\
$m_{\pi\pi^0}$ Dependence ($\sigma_{\rm \lambda}$) & $-$ & $-$ & $-$ & $-$ & 0.16 &0.16 \\
\hline
Total ($\sigma_{\rm tot}$)        & 0.66 & 1.17 & 0.97 & 1.53 & 0.94 & 1.81 \\ \hline
\end{tabular}
\end{center}
\end{table}

The impact of a possible systematic mismeasurement of $\deltat$ ($\sigma_{\Delta t}$) has been estimated by comparing different
parameterizations of  the resolution function, varying the position of the beam spot and the absolute $z$ scale within their uncertainties, 
and loosening and tightening the quality criteria on the reconstructed decay points. 
We also estimate the impact of the uncertainties on the alignment of the silicon vertex tracker (SVT) by repeating the measurement using simulated events, with the SVT intentionally misaligned.
For the systematic uncertainty of the fit ($\sigma_{\rm fit}$), we quote the upper limit on the bias on the $a^{\mu}$ and $c^{\mu}$ parameters,  
as estimated from samples of fully-simulated events. 
The model error ($\sigma_{\rm mod}$) contains the uncertainty on the $B^{0}$ lifetime and \deltamd,
varied by the uncertainties on the world averages~\cite{PDG} and also by allowing them to vary in the fit. 
The tagging error ($\sigma_{\rm tag}$) is estimated considering possible differences in tagging efficiency between \Bz\ and \Bzb,
different mistag fractions for the decay modes $D\pi,\, D^{*}\pi,\, D\rho$,
and different $\Delta t$ resolutions for correctly and incorrectly tagged events. 
We also account for uncertainties in the background ($\sigma_{\rm bkg}$) by varying the effective lifetimes, dilutions, \mes~shape parameters, signal fractions, and background \CP asymmetry.
A possible dependence of $a^{D\rho}$ $(c_{\rm lep}^{D\rho})$ on the $\pi\pi^0$ invariant mass ($\sigma_{\rm \lambda}$) is estimated from the
limit on the fraction of $\Bz {\to} D^-\rhop(1450)$ and $\Bz {\to} D^-(\pi^+\piz)_{nr}$ in the $\rho$ mass window.

As a cross-check, we perform the same fits on a sample of $6843$ $\Bm{\to} D^{(*)0}\pim$ candidates,
where, as expected, we find no \CP\ asymmetries.  
We combine our results with the result obtained on the partially reconstructed $B {\to} D^{*\pm}\pi^{\mp}$ sample~\cite{partial} 
and use a frequentist method described in Ref.~\cite{partial} to set a constraint on $2\beta\!+\!\gamma$. 
The confidence level as a function of $|\stwobg|$ is shown in Figure~\ref{fig:s2bg1}. 
We set the lower limits $|\stwobg|\!>\!0.64\ (0.40)$ at $68\%$ $(90\%)$ C.L.

\begin{figure}[!htb]
\begin{center}
\epsfig{figure=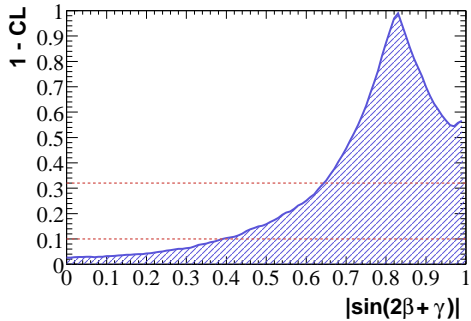,height=4.24cm,clip=}
\caption{Frequentist confidence level as a function of $|\stwobg|$,
obtained when combining our result with the result obtained on
partially reconstructed $B {\to} D^{*\pm}\pi^{\mp}$ decays~\cite{partial}.
The horizontal lines show the $68\%$ (top) and $90\%$ C.L. (bottom).} \label{fig:s2bg1}
\end{center}
\end{figure}
%

In conclusion, we have studied the time-dependent \CP-violating asymmetries in fully reconstructed $\Bz{\to}D^{\pm}\pi^{\mp}$, $\Bz{\to}D^{*\pm}\pi^{\mp}$, and $\Bz{\to}D^{\pm}\rho^{\mp}$ decays
in a sample of $232$ million $\Y4S{\to}\BB$ decays, and have measured the \CP-violating parameters listed in Eq.~\ref{acmeas}.
We interpret the result in terms of $\stwobg$ and find that $|\stwobg|\!>\!0.64$ $(0.40)$ at $68\%$ $(90\%)$ C.L.
These results are consistent with and supersede our previous measurement.

We are grateful for the excellent luminosity and machine conditions
provided by our \pep2\ colleagues, 
and for the substantial dedicated effort from
the computing organizations that support \babar.
The collaborating institutions wish to thank 
SLAC for its support and kind hospitality. 
This work is supported by
DOE
and NSF (USA),
NSERC (Canada),
IHEP (China),
CEA and
CNRS-IN2P3
(France),
BMBF and DFG
(Germany),
INFN (Italy),
FOM (The Netherlands),
NFR (Norway),
MIST (Russia), and
PPARC (United Kingdom). 
Individuals have received support from CONACyT (Mexico), A.~P.~Sloan Foundation, 
Research Corporation,
and Alexander von Humboldt Foundation.


\begin{thebibliography}{99}
\bibitem{chconj}
Charge conjugation is implied in this paper, unless otherwise stated. 
The superscript $(*)$ indicates that the superscripted symbol must be considered both with and without the $*$.
\bibitem{note}
According to Ref.~\cite{fleischer}, the  strong 
phase difference for $\Bz{\to} D^{*\pm} \pi^{\mp}$ is given by $\delta^{D^*\pi} + \pi$, but this does not affect this measurement.
\end{thebibliography}
\end{document}